\documentclass[conference]{IEEEtran}
\IEEEoverridecommandlockouts
\usepackage{cite}
\usepackage{amsmath,amssymb,amsfonts}
\usepackage{bm}
\usepackage{algorithm}
\usepackage{algorithmic}

\usepackage{graphicx}
\usepackage{textcomp}
\usepackage{xcolor}

\def\BibTeX{{\rm B\kern-.05em{\sc i\kern-.025em b}\kern-.08em
    T\kern-.1667em\lower.7ex\hbox{E}\kern-.125emX}}

\newtheorem{lemma}{Lemma}

\begin{document}
\title{
Two-Stage Prony-Based Estimation of Fractional Delay and Doppler Shifts in OTFS Modulation
\thanks{This work was supported in part by JSPS KAKENHI Number JP23K26104 and JP23H00474.}
}

\author{
\IEEEauthorblockN{Yutaka Jitsumatsu}
\IEEEauthorblockA{\textit{Deptartment of Informatics,} \\
\textit{Kyushu University, Japan}\\
jitsumatsu@inf.kyushu-u.ac.jp}

\and
\IEEEauthorblockN{Liangchen Sun}
\IEEEauthorblockA{\textit{Department of Information Science and Technology} \\
\textit{Kyushu University, Japan}\\
sun@me.inf.kyushu-u.ac.jp}
}

\maketitle

\begin{abstract}
This paper addresses the estimation of fractional delay and Doppler shifts in multipath channels that cause doubly selective fading—an essential task for integrated sensing and communication (ISAC) systems in high-mobility environments. Orthogonal Time Frequency Space (OTFS) modulation enables simple and robust channel compensation under such conditions. However, fractional delay and Doppler components introduce inter-path interference, degrading estimation accuracy.
We propose a two-stage estimation method based on Prony’s technique using OTFS pilot signals with $M$ subchannels and $N$ pilot repetitions. In the first stage, Doppler frequencies are estimated by jointly solving $M$ coupled Prony equations, exploiting the periodicity of the pilot signal. In the second stage, delays are estimated by applying the discrete Fourier transform (DFT) and Prony’s method to each Doppler component obtained in the first stage.
The proposed method can accurately estimate up to $N-1$ delay-Doppler parameters under noiseless conditions. In noisy environments, conventional information criteria such as AIC and BIC yield suboptimal performance; thus, a heuristic model order selection is adopted. Numerical simulations confirm that the proposed method achieves high estimation accuracy, highlighting its potential for future ISAC frameworks.
\end{abstract}


\section{Introduction}

Orthogonal Time Frequency Space (OTFS)\cite{OTFS, OTFS-BITS, OTFS-BITS2} modulation has gained increasing attention as a promising solution for high-mobility communication systems, thanks to its robustness against Doppler spread.
In addition to communications, OTFS is also seen as a strong candidate for integrated sensing and communication (ISAC) systems\cite{yuan2021integrated, shi2023integrated, zegrar2024otfs}, where radar and communication functionalities are jointly realized within a single framework.
Accurate estimation of delay and Doppler parameters in multipath channels is a crucial step for both applications.
When these parameters lie exactly on the integer grid in the delay-Doppler (DD) domain, OTFS allows simple and effective estimation.
However, in practical environments, delays and Doppler shifts often contain fractional components. These result in energy leakage across neighboring DD bins and mutual interference among multipath components, significantly degrading estimation performance.

Several methods have been proposed to tackle this challenge. Gaudio et al.~\cite{gaudio2020effectiveness} demonstrated the potential of OTFS for joint radar parameter estimation and communication. More recent studies have addressed fractional parameter estimation: Muppaneni et al.~\cite{muppaneni2023channel} proposed a delay-Doppler estimation technique for fractional values based on pilot correlation; Zhang et al.~\cite{zhang2023radar} presented a signal-processing-oriented view of OTFS for radar sensing; Zacharia and Devi~\cite{zacharia2023fractional} analyzed fractional parameter estimation performance in ISAC scenarios; and Ranasinghe et al.~\cite{ranasinghe2024fast} introduced a sequential method for MIMO-OTFS systems. Despite these advancements, many existing approaches either rely on grid-based methods or require a large number of pilots, which limit estimation precision or efficiency, especially at high SNR.

In this paper, we propose a two-stage estimation method based on Prony's technique for the accurate extraction of fractional delay and Doppler parameters from OTFS pilot signals. Prony's method is well known for its high resolution in estimating the frequencies of complex exponentials in low-noise settings. To apply this one-dimensional technique to two-dimensional delay-Doppler estimation, we introduce the following three key innovations:

First, we design the pilot signal to be DFT-friendly, ensuring that the frequency corresponding to $m = -M/2$ is correctly handled when interpreting the DFT from $m = 0, 1, \ldots, M-1$. This correction is essential, as the periodicity of the signal is disturbed in the presence of fractional Doppler shifts.

Second, we arrange the received signal as a two-dimensional matrix indexed by slow and fast time, such that Doppler shifts can be isolated along the slow-time axis. This structure motivates a two-stage approach: Doppler frequencies are estimated first, and then the corresponding delay components are extracted after Doppler-induced effects are removed.

Third, we develop a novel formulation of coupled Prony equations based on the periodicity of the pilot signal. This allows us to jointly solve multiple equations with shared solutions, enhancing estimation robustness and increasing the maximum number of resolvable paths compared to the standard Prony method.

We consider a communication system based on OTFS modulation. We assume that frame synchronization has already been established. While frame synchronization is an important prerequisite for initiating communication, it is beyond the scope of this work. Since the channel conditions may vary during communication, it is necessary to update the channel estimates on a block-by-block basis.

The remainder of this paper is organized as follows. Section II introduces the system model and outlines the structure of the transmitted and received signals. Section III presents the proposed two-stage estimation method. Section IV provides simulation results and discusses model order selection via information-theoretic criteria. Section V concludes the paper.

\section{Problem Formulation}
\begin{figure}[t]
    \centering
    \includegraphics[width=0.95\linewidth]{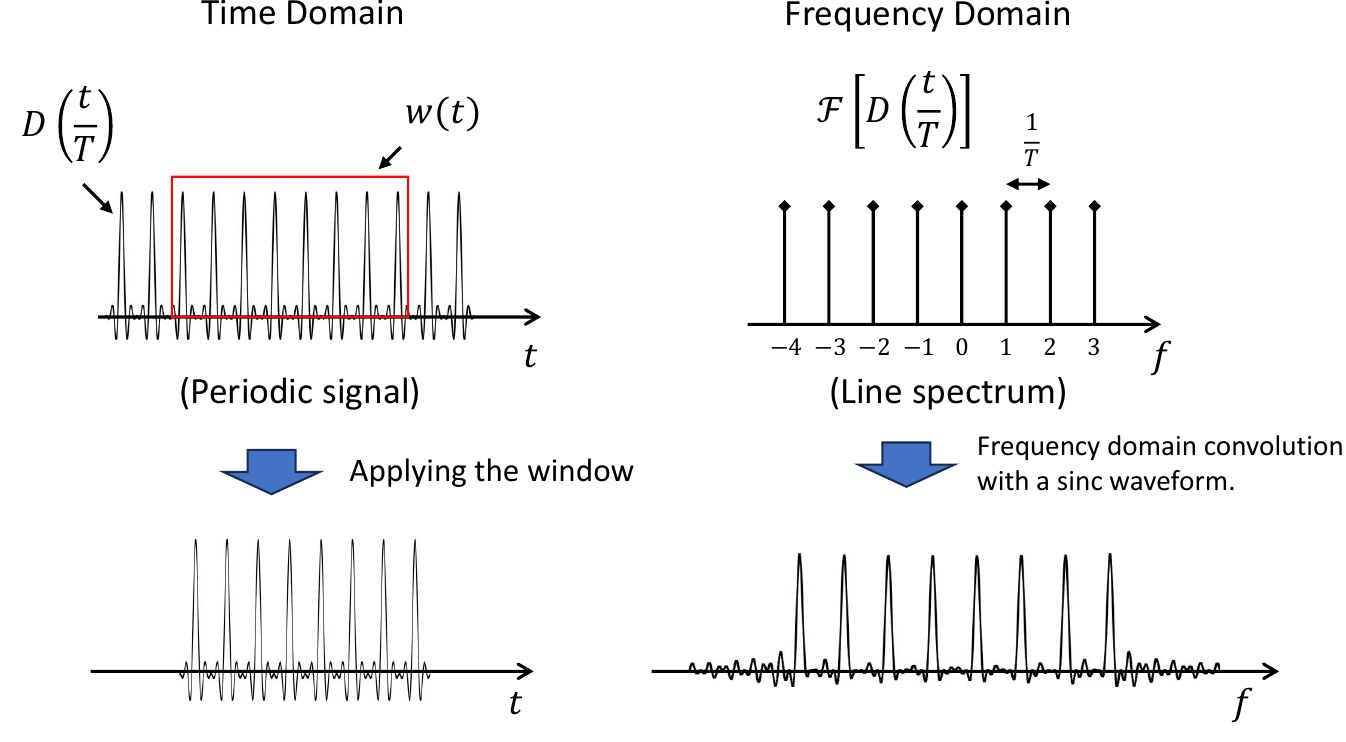}
    \caption{Construction of the proposed transmittted signal, where $N=M=8$.}
    \label{fig:construction_transmitted_signal}
\end{figure}

We use a center-shifted variant of the Dirichlet kernel as the transmitted pilot signal for delay--Doppler estimation. The signal is defined as
\begin{align}
s(t) &= D_M\left( \frac{t}{T} \right) w(t),
\label{s(t)}\\
D_M(x) &= \sum_{m = -M/2}^{M/2 - 1} e^{j 2\pi m x} = e^{-j \pi x} \cdot \frac{\sin(\pi M x)}{\sin(\pi x)},
\label{D_M(x)}
\end{align}
where $T$ denotes the duration of a time slot and $w(t)$ is a rectangular window function of duration $NT$,
with $N$ being the number of pilot repetitions.  
The signal construction is illustrated in Figure~\ref{fig:construction_transmitted_signal}.
In the frequency domain, $D_M(\frac{t}{T})$ has line spectrums
at $f=\frac{m}{T}$ for $m=-M/2, \ldots, M/2-1$.

Note that Eq.~(\ref{D_M(x)}) defines a periodic signal with period one.  
This choice offers a significant advantage: the reception process and delay--Doppler estimation can be efficiently performed using the DFT.  
The window function $w(t)$ is chosen to shift the center such that the time-domain signal is truncated between its peaks. This helps suppress power leakage in the frequency domain.

This signal is similar to, but distinct from, the standard OTFS pilot signal, which is defined in the delay--Doppler (DD) domain as
\begin{align}
X_\mathrm{DD}[k,\ell] = \begin{cases}
1, & k = \ell = 0, \\
0, & \text{otherwise}.
\end{cases}
\label{pilot-DDdomain}
\end{align}
A detailed comparison is provided in Appendix~\ref{appendix:pilotsignal}.

Consider a multipath channel with Doppler shift. Let $P$ denote the number of propagation paths, which is unknown to the receiver. However, we assume that the channel is \textit{sparse}, in the sense that $P$ is significantly smaller than $NM$, the approximate dimension of the signal space.

Let $t_{d,\max}$ and $f_{D,\max}$ denote the maximum delay and maximum Doppler shift, respectively.  
For the $p$-th path, let $\alpha_p$, $t_{d,p}$, and $f_{D,p}$ denote the attenuation factor, time delay, and Doppler shift.  
We assume that $\alpha_p$ follows a complex Gaussian distribution. The time slot duration $T$ must be chosen such that $t_{d,p}$ and $f_{D,p}$ satisfy
\[
0 < t_{d,p} < T, \quad -\frac{1}{2T} < f_{D,p} < \frac{1}{2T}.
\]
This implies the following relationship:
\[
t_{d,\max} \le T \le \frac{1}{2f_{D,\max}}.
\]

The received signal is defined as
\begin{align}
    r(t) = \sum_{p=1}^P \alpha_p s(t - t_{d,p}) e^{j 2\pi f_{D,p} t} + z(t),
    \label{r(t)}
\end{align}
where $z(t)$ denotes additive white Gaussian noise (AWGN).  
The received signal is sampled with a sampling interval of $T_s = T/M$.  
The discrete-time samples are given by
\begin{align}
    r_\mathrm{TD}[\ell] 
    = \int r(t)\, \mathrm{sinc}\left( \frac{t}{T_s} - \ell \right) dt \approx r(\ell T_s),
\end{align}
where $\mathrm{sinc}(x) = \frac{\sin(\pi x)}{\pi x}$.  
The approximation holds under the assumption that the maximum Doppler shift $f_{D,\max} < 1/T$ is much smaller than the bandwidth of the transmitted signal, which is approximately $1/T_s=M/T$.

\begin{lemma}
Let $\bm{R} = ( R_{n,\ell} )$ be an $N\times M$ matrix defined by
\[
R_{n,\ell} = r_{\rm TD}[nM+\ell],
\]
and define the $N \times P$ matrix $\bm{E} = (E_{n, p})$ by
\[
E_{n, p} = e^{j2\pi f_{D,p} nT}.
\]
Also, let $\bm{V} = (V_{p,\ell})$ be a $P \times M$ matrix whose $(p,\ell)$-component is given by
\[
V_{p,\ell} = \alpha_p D_M\left( \frac{\ell}{M} - \frac{t_{d,p}}{T} \right) e^{j2\pi f_{D,p} \ell T_s}.
\]
Then, we have
\begin{align}
\bm{R} = \bm{E} \bm{V}.
\label{R=EV}
\end{align}
\end{lemma}
The proof is omitted because of the space limitation.

This lemma suggests that the Doppler frequencies $f_{D,p}$ can be estimated based on the fact that the matrix $\bm{E}$ is determined solely by $\{ f_{D,p} \}$, regardless of $\{ t_{d,p} \}$.
Such a decomposition leads to a two-stage approach: we first estimate all $f_{D,p}$, and then estimate the corresponding $t_{d,p}$ for each $f_{D,p}$.
The attenuation factor $\alpha_p$ can also be estimated in the second stage.

Note that in a standard OTFS demodulation, the $N$-point discrete Fourier transform (DFT) of $R_{n,\ell}$ with respect to $n$ is performed to obtain the DD domain received signal $Y_\mathrm{DD}[k, \ell]$, i.e.
$Y_\mathrm{DD}[k, \ell] = \bm{F}_N \bm{R}$, where $(\bm{F}_N)_{k,n} = e^{-j \frac{2 \pi}{N} kn }$.
Let
\begin{align}
    t_{d,p} = (\ell_p + \epsilon_{t,p})T_s, \quad
    f_{D,p} = \frac{k_p+\epsilon_{f,p} }{NT},
\end{align}
where $\ell_p$ and $k_p$ are integer parts and $\epsilon_{t,p}$ and $\epsilon_{f,p} $ 
are fractional parts of the delay and the Doppler. 
The expression (\ref{R=EV}) suggests that if $f_{D,p}$ does not have a fractional part,
then the matrix $\bm{E}$ consists of columns that are aligned with the $N$-point DFT basis vectors.
As a result, the transformation $\bm{F}_N \bm{E}$ yields a sparse matrix whose nonzero entries are concentrated at specific Doppler indices.
Consequently, the DD domain signal $Y_\mathrm{DD}[k, \ell] = (\bm{F}_N \bm{R})_{k,\ell}$ exhibits sparsity along the Doppler dimension,
which facilitates efficient detection and estimation of the channel parameters.

In practice, however, the Doppler frequency $f_{D,p}$ typically includes a nonzero fractional part, which causes energy leakage across multiple Doppler bins and thus limits the sparsity of $Y_\mathrm{DD}[k,\ell]$.
In such cases, parametric estimation methods such as Prony’s method are effective in accurately recovering the channel parameters.

Suppose that the Doppler shifts have been perfectly estimated. This implies that the matrix $V$ has been obtained. 
Next, we remove the Doppler-induced phase components from each element $V_{p,\ell}$. 
After this compensation, we obtain the modified signal
\begin{align}
    \widetilde V_{p,\ell} = \alpha_p D_M\left( \frac{\ell}{M} - \frac{t_{d,p}}{T} \right).
\end{align}
%
The $M$-point DFT of $\widetilde V_{p,\ell}$, $\ell = 0,1,\ldots, M-1$ is given by
\begin{align}
& \sum_{\ell =0}^{M-1} \widetilde V_{p,\ell} \, e^{-j\frac{2\pi}M m\ell} \notag\\
& = \sum_{\ell =0}^{M-1} \alpha_p D_M\left( \frac{\ell}{M} - \frac{t_{d,p}}{T} \right) e^{-j\frac{2\pi}M m\ell} \notag\\ 
& = \alpha_p  \sum_{m'=-\frac{M}{2}}^{\frac{M}{2}-1} \sum_{\ell =0}^{M-1} e^{j2\pi m' \left( \frac{\ell}{M} - \frac{t_{d,p}}{T} \right)} e^{-j\frac{2\pi}M m\ell} \notag\\
& = \alpha_p M  e^{- j2\pi m' \frac{t_{d,p}}{T} } ,
\label{DFT_of_V_tilde}
\end{align}
where $m' = m$ for $0 \le m \le \frac{M}{2}-1$, and $m' = m - M$ for $\frac{M}{2} \le m \le M - 1$.
This result suggests that $t_{d,p}$ can be estimated by applying Prony's method to the vector $\bm{\widetilde{V}}$.

\section{Two-Stage Prony Method}
This section presents the proposed Prony-based method, consisting of two estimation stages.

\subsection{Stage 1: Doppler Estimation}
The first stage estimates the Doppler shifts. Since $\bm{R}$ has $M$ columns, Prony’s method is applied to each column individually.  
All resulting Prony equations share the same set of solutions, which correspond to the Doppler frequencies.  
This allows us to formulate $M$ simultaneous equations for estimating the Doppler shifts more robustly.  
To the best of the authors’ knowledge, such an approach has not been reported in the literature.

Let $\hat{P}$ denote the prescribed number of paths.
Its value is determined based on an information criterion.
The maximum number of $\hat{P}$ is $N-1$, which is determined by (\ref{T^l}) below. 
Selecting an appropriate value for $\hat{P}$ is one of the most challenging aspects of the proposed method.
This issue will be discussed in Section~\ref{section:MdelOrderSelection}.

For each $\ell = 0, 1, \ldots, M-1$, construct a Toeplitz matrix $T^{(\ell)}$ as
\begin{align}
    T^{(\ell)}
    &=
    \begin{pmatrix}
        R_{\hat P, \ell} & R_{\hat P-1, \ell} & \cdots & R_{0, \ell} \\
        R_{\hat P+1, \ell} & R_{\hat P, \ell} & \cdots & R_{1, \ell} \\
        \vdots             & \vdots       &        & \vdots\\
        R_{N-1, \ell} & R_{N-2, \ell} & \cdots & R_{N-\hat P-1, \ell} 
    \end{pmatrix}.
    \label{T^l}
\end{align}
Then, stack these matrices vertically to form a merged matrix $T$ as
\begin{align}
    T = \begin{bmatrix}
        T^{(0)}\\
        \vdots\\
        T^{(M-1)}
    \end{bmatrix}.
\end{align}

The Prony method first determines a vector $\bm{a} = (a[0], a[1], \ldots, a[\hat{P}])^t$ with $a[0]=1$ that satisfies
$\bm{T} \bm{a} = \bm{0}$, where $\bm{0}$ is a zero vector of dimension $M \times (N - \hat{P})$.
Let $\bm{t}_0$ denote the first column of $\bm{T}$, and let $\widetilde{\bm{T}}$ denote the matrix obtained by removing the first column from $\bm{T}$. Then, we obtain
\begin{align}
\bm{a}^t &= (1, \tilde{\bm{a}}^t), \\
\tilde{\bm{a}} &= -\widetilde{\bm{T}}^\dagger \bm{t}_0,
\end{align}
where $\widetilde{\bm{T}}^\dagger$ is a generalized inverse of $\widetilde{\bm{T}}$.

Then, find the zeros of the polynomial
\begin{align}
a[0] x^{\hat{P}} + a[1] x^{\hat{P} -1} + \cdots + a[\hat{P} -1] x + a[\hat{P}] = 0.
\label{polynomial}
\end{align}
Denote them by $Z_p$, for $p = 1, 2, \ldots, \hat{P}$.
The estimated Doppler frequency is given by
\begin{align}
\hat{f}_{D,p} = \frac{\arg(Z_{p})}{2\pi T},
\end{align}
where $\arg(Z_p)$ denotes the argument (i.e., phase angle) of the complex number $Z_p$.

\subsection{Preprocessing before Stage 2}
Using the estimated Doppler shifts $\hat{f}_{D,p}$, 
we reconstruct the matrix $\bm{\hat{E}}$ as 
\begin{align}
    \bm{\hat{E}} = \left( e^{j2\pi \hat{f}_{D, p} n T} \right)_{n,p}.
\end{align}
Then, from (\ref{R=EV}), the matrix $\bm{V}$ is estimated as
\begin{align}
    \hat{\bm{V}} 
    = \arg \min_{ \bm{V} } \left\| \bm{R} - \bm{\hat{E}} \bm{V} \right\|^2 
    = \bm{\hat{E}}^{\dagger} \bm{R},
\end{align}
where $\dagger$ denotes the Moore–Penrose pseudo-inverse.
To cancel the Doppler effect in $\hat{\bm{V}}$, we compute  
\begin{align}
    \widetilde{V}_{p,\ell} = \hat{V}_{p,\ell} \, e^{ -j2\pi \hat{f}_{D,p} \ell T_s}. 
\end{align}

Next, we apply the $M$-point DFT to $\widetilde{V}_{p,\ell}$ for each $p$ to obtain
$Y_p[m] = \sum_{\ell=0}^{M-1} \widetilde{V}_{p,\ell} \, W_M^{m\ell}$,
where $W_M = e^{-j \frac{2\pi}{M}}$ is the $M$-th root of unity.
If the Doppler shift estimates $\hat{f}_{D,p}$ are perfect and noise is absent, it follows from (\ref{DFT_of_V_tilde}) that
\begin{align}
    Y_p[m] = \alpha_p M \, e^{-j2\pi m t_{d,p} / T}
\end{align}
holds for $m = -\frac{M}{2}, -\frac{M}{2} + 1, \ldots, \frac{M}{2} - 1$.

{\bf Remark}:
Since $Y_p[m]$ is not periodic, the index range for $m$ must be carefully chosen.
While for periodic signals, the DFT is invariant under circular shifts of the index (e.g., $[0, \dots, M-1]$ vs. $[-M/2, \dots, M/2 - 1]$), this equivalence no longer holds for non-periodic complex exponentials.
As a result, using the symmetric interval centered around zero is essential for correct interpretation.

\subsection{Stage 2: Delay and Attenuation Estimation}

Let $L$ denote the number of paths sharing the same Doppler shift. The selection of $L$ will be discussed in Section~\ref{section:MdelOrderSelection}.

For each vector $\bm{Y}_p = (Y_p[m])_m$, where $p = 1, 2, \ldots, \hat{P}$, we apply Prony's method by constructing a Toeplitz matrix:
\begin{align}
    (\bm{T}')_{ij} = Y_p\left[L - \textstyle\frac{M}{2} + i - j\right],
\end{align}
where $i = 1, \ldots, M - L$ and $j = 1, \ldots, L + 1$.

We then find a nonzero vector $\bm{a} = (a[0], a[1], \ldots, a[L])^\mathrm{T}$ satisfying $\bm{T}' \bm{a} = \bm{0}$, and compute the roots of the polynomial

(\ref{polynomial}) where $\hat{P}$ in (\ref{polynomial}) is replaced by $L$. Denote the roots of this polynomial by $Z_{p,\ell}$.

The estimated time delays corresponding to the Doppler shift $\hat{f}_{D,p}$ are then given by
\begin{align}
    \hat{t}_{d,p,\ell} = \frac{\arg Z_{p,\ell}}{2\pi} T,
\end{align}
for $\ell = 1, 2, \ldots, L$.

The determination of the estimated number of distinct Doppler shifts, denoted by \( \hat{P} \), and the number of delays \( L \) associated with each Doppler shift, is of crucial importance.

Figure~\ref{fig:residual_vs_P_hat} shows an example of the residual sum of squares (RSS), defined as
\begin{align}
    \mathrm{RSS} = \min_{\lVert \bm{a} \rVert^2 = 1} \lVert \bm{T} \bm{a} \rVert^2,
\end{align}
for different values of \( \hat{P} \) in the absence of noise. The parameters are set as \( N = M = 16 \) and \( P = 5 \). The maximum delay and Doppler shift are \( t_{d,\max} = T \) and \( f_{d,\max} = 1/(2T) \), respectively. The path gains satisfy \( |\alpha_p| = 1 \) for all \( p \), and the values of \( t_{d,p} \), \( f_{D,p} \), and the phase angles of \( \alpha_p \) are independently and uniformly distributed.
As shown in the figure, when \( \hat{P} \geq 5 \), the residual error becomes nearly zero.

\subsection{Model order selection}
\label{section:MdelOrderSelection}
\begin{figure}
    \centering
    \includegraphics[width=0.85\linewidth]{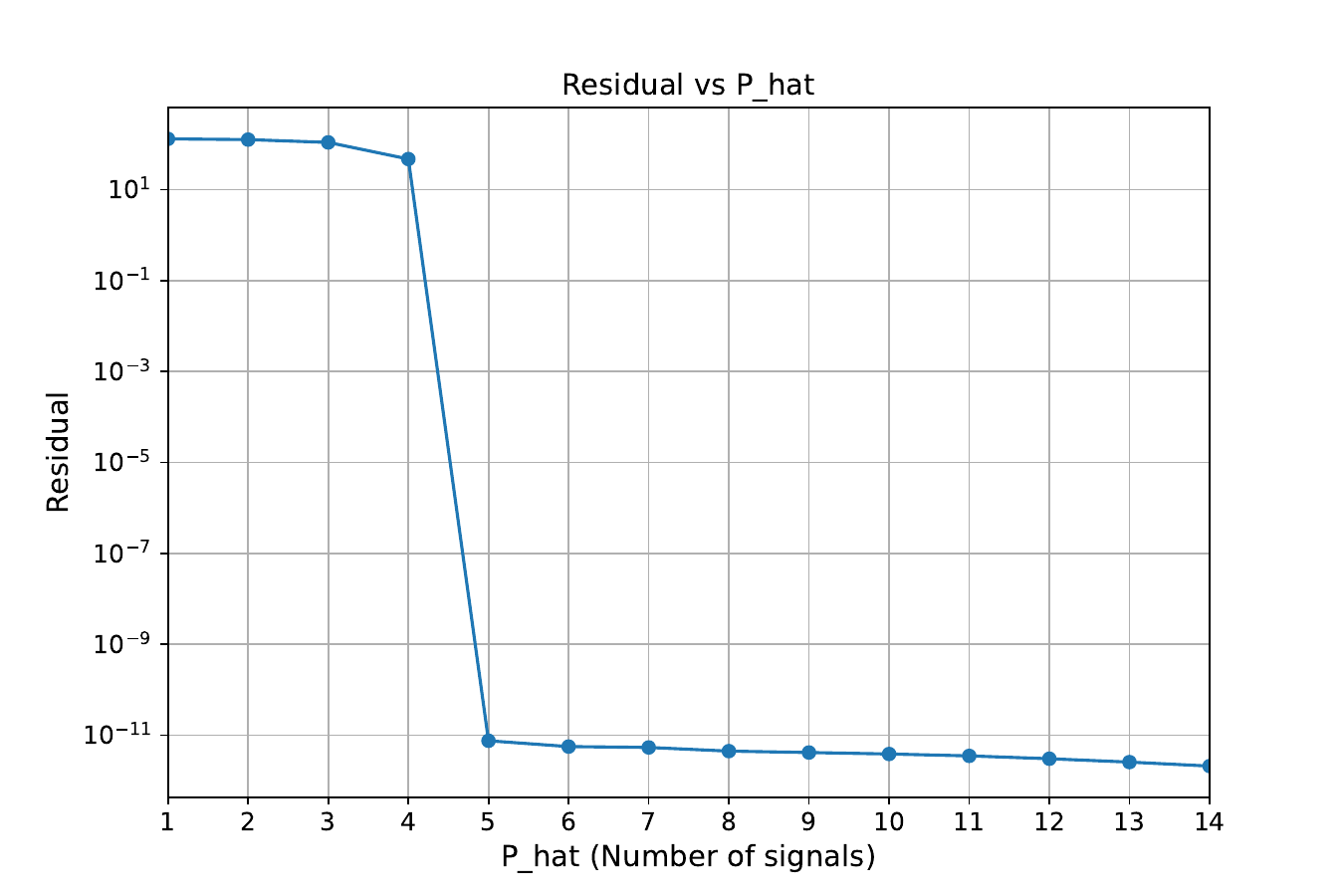}
    \caption{The residual error of $\lvert \bm{Ta} \rVert^2$ against $\hat P$. Noiseless case.}
    \label{fig:residual_vs_P_hat}
\end{figure}
\begin{figure}
    \centering
    \includegraphics[width=0.85\linewidth]{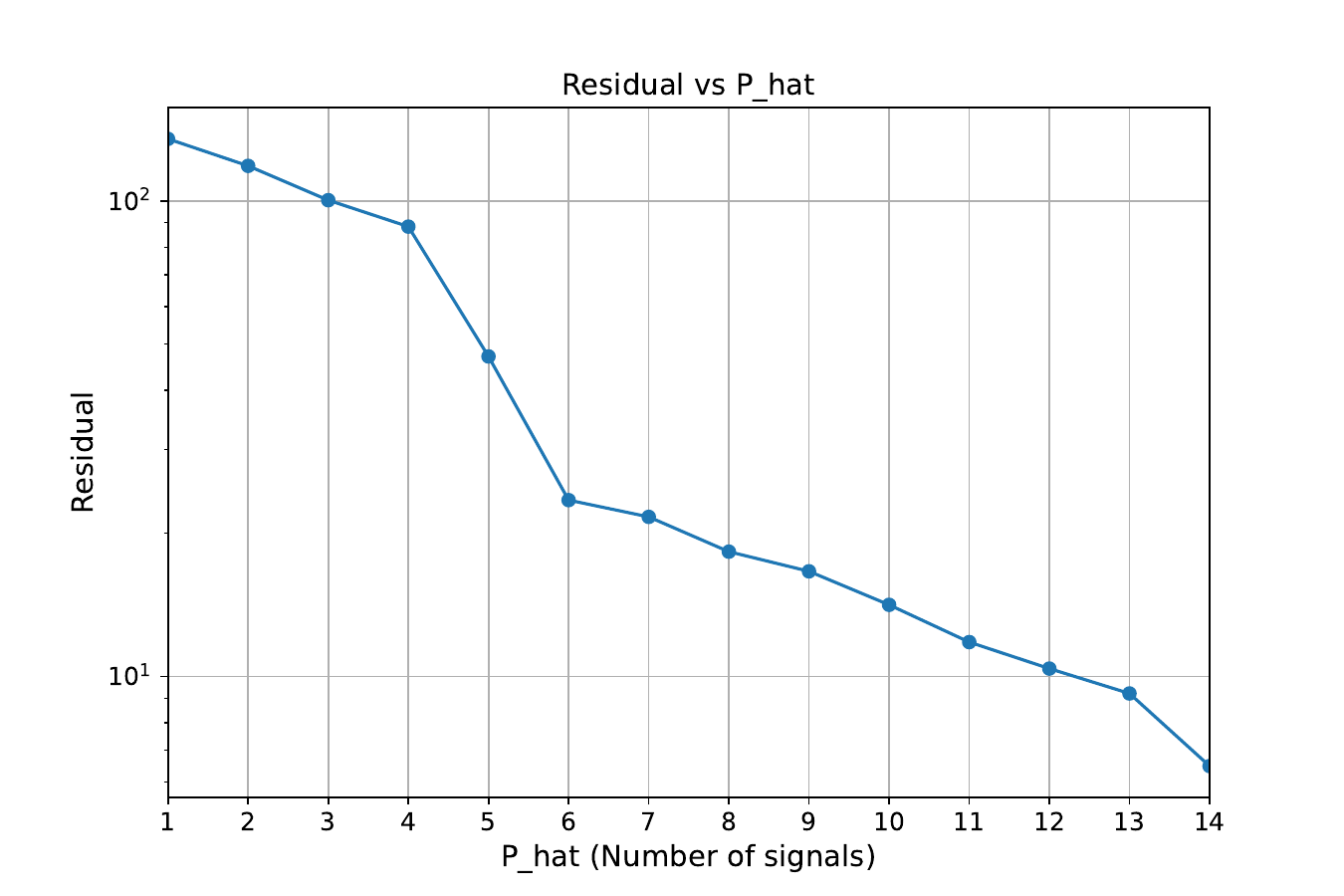}
    \caption{The residual error of $\lvert \bm{Ta} \rVert^2$ against $\hat P$, where SNR is $20$ dB.}
    \label{fig:residual_vs_P_hat20dB}
\end{figure}

Figure~\ref{fig:residual_vs_P_hat20dB} presents the RSS curve for a noisy case with SNR = 20 dB. The RSS decreases monotonically as \( \hat{P} \) increases. However, a noticeable change in the slope is observed: the rate of decrease becomes slower for \( \hat{P} > P \) when the RSS is plotted on a logarithmic scale.

To determine the optimal value of $\hat{P}$ at Stage 1, an information-theoretic criterion, such as Akaike’s Information Criterion (AIC) or the Bayesian Information Criterion (BIC), should be employed.
Specifically, AIC and BIC are defined as
\begin{align}
\mathrm{AIC} &= 2 (3 \hat{P} ) + N \log \left ( \frac{\mathrm{RRS}}{N} \right) + \mathrm{const}, \\
\mathrm{BIC} &= (3 \hat{P} ) \log N + N \log \left ( \frac{\mathrm{RRS}}{N} \right) + \mathrm{const}.
\end{align}
$(3\hat{P})$ means the number of parameters $(\alpha_p, t_{d,p}, f_{D,p})$. $|\alpha_p|$ was set to one only for drawing Figure~\ref{fig:residual_vs_P_hat}. For simulation $\alpha_p$ follows a complex Gaussian distribution.  
The choice of an appropriate information criterion is a common issue in hyper-parameter estimation problems.

A key challenge in the proposed method is accurately estimating the number of propagation paths, \( P \).  
Although model selection criteria such as AIC and BIC are fundamental tools for this purpose, they often fail to provide reliable estimates in practice.  
These criteria tend to either overestimate or underestimate the true value, making it difficult to determine \( P \) robustly.  
Heuristic approaches may therefore be helpful in addressing this issue.

Since AIC and BIC do not provide reliable estimates in our setting, we employed the following heuristic approach instead.  
Let \( \hat{P} \) be a temporary overestimate of the true number of paths \( P \).  
Stage~1 is performed to obtain Doppler frequency estimates \( \hat{f}_{D,p} \).  
Using these estimates, the preprocessing step before Stage~2 is executed to construct \( \bm{\tilde{V}} \).  
Let the squared \( L^2 \)-norm of the \( p \)-th row of \( \bm{\tilde{V}} \) be denoted by \( \lVert \bm{V}_p \rVert^2 \).  
Then, all rows with squared \( L^2 \)-norms less than 10\% of the maximum among all rows are discarded.

For delay estimation in Stage~2, we employed AIC and BIC, but found it challenging to determine whether the number of delays is one or greater than one. Therefore, we set \( L = 1 \) throughout our simulations. This assumption limits the capability of the proposed method when multiple paths share nearly identical Doppler shifts.

Further investigation is needed to determine an appropriate value of \( L \).  
However, in high SNR scenarios, it is still possible to distinguish Doppler frequencies \( f_{D,p} \) that are very close to each other.  
In such cases, the assumption \( L = 1 \) remains valid.

\section{Simulation results}

\begin{figure}[t]
    \centering
    \includegraphics[width=0.85\linewidth]{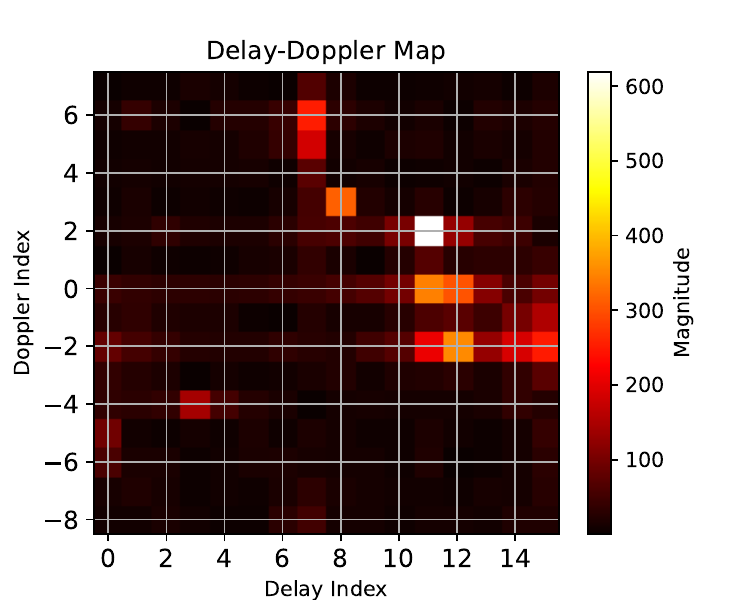}
    \caption{The absolute value of $Y_\mathrm{DD}[k,\ell]$, where $N=M=16$, $P=8$, and SNR is $20$dB}
    \label{fig:absolute_value_reveived_signal}
\end{figure}

\begin{figure}[t]
    \centering
    \includegraphics[width=0.85\linewidth]{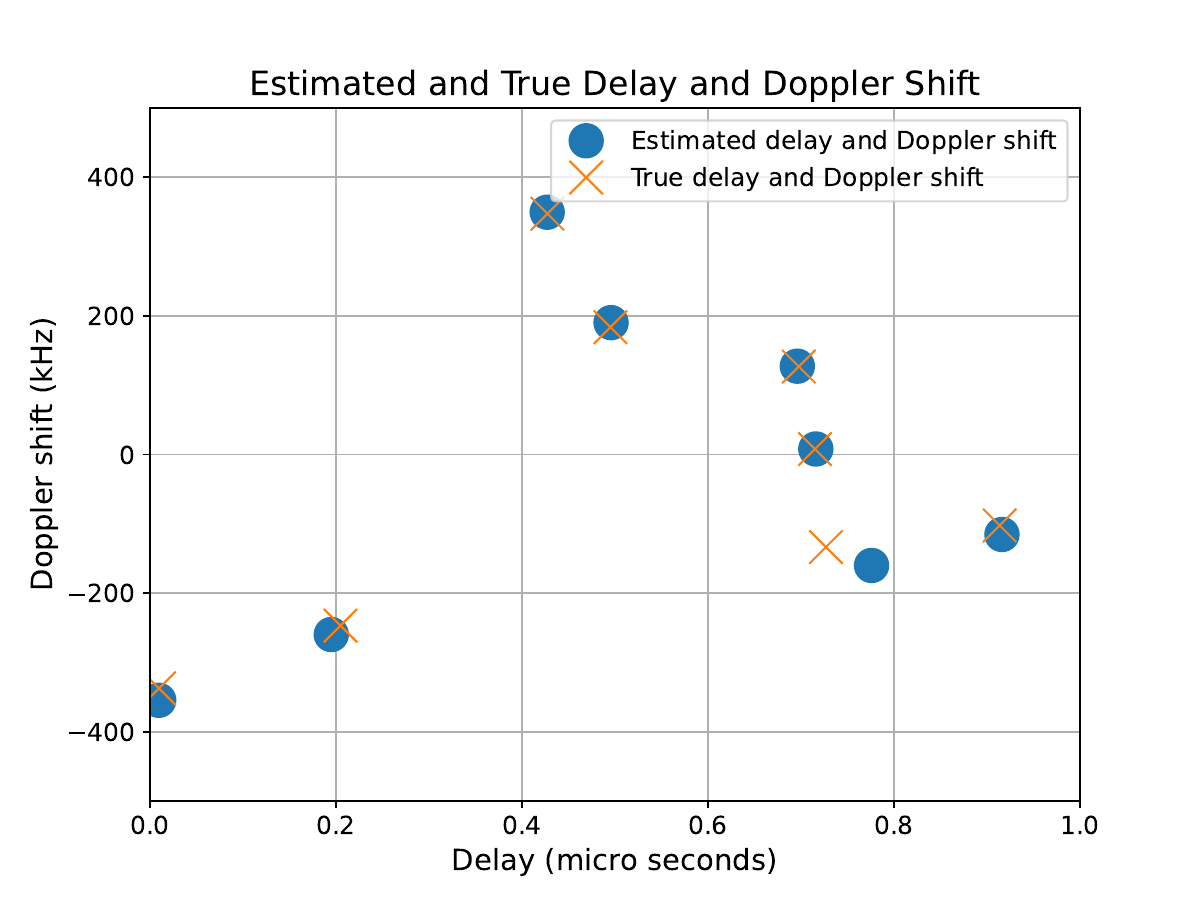}
    \caption{The true and the estimated delay and Doppler values for the same example shown in Fig.~\ref{fig:absolute_value_reveived_signal}}
    \label{fig:estimated_delay_DopplerN=M=16P=10}
\end{figure}

Numerical simulation results are reported in this section.  
Let \( T = 1.0 \times 10^{-6} \) seconds.  
Figure~\ref{fig:absolute_value_reveived_signal} shows the absolute value of the received signal \( R_{n,\ell} = r_{\rm TD}[nM+\ell] \), for \( N = M = 16 \), \( P = 8 \), and SNR = 20~dB.  
Figure~\ref{fig:estimated_delay_DopplerN=M=16P=10} presents the true and estimated delay–Doppler values obtained by the proposed method, which shows high accuracy. In this example, AIC was used to estimate \( \hat{P} \).  
From the \( Y_{\rm DD}[k,\ell] \) map, the integer parts of the delay and Doppler shifts can be inferred, though energy leakage may obscure the peaks.  
Conventional methods refine the fractional parts after coarse estimates~\cite{zhang2023radar}.  


\begin{figure}[t]
    \centering
    \includegraphics[width=0.95\linewidth]{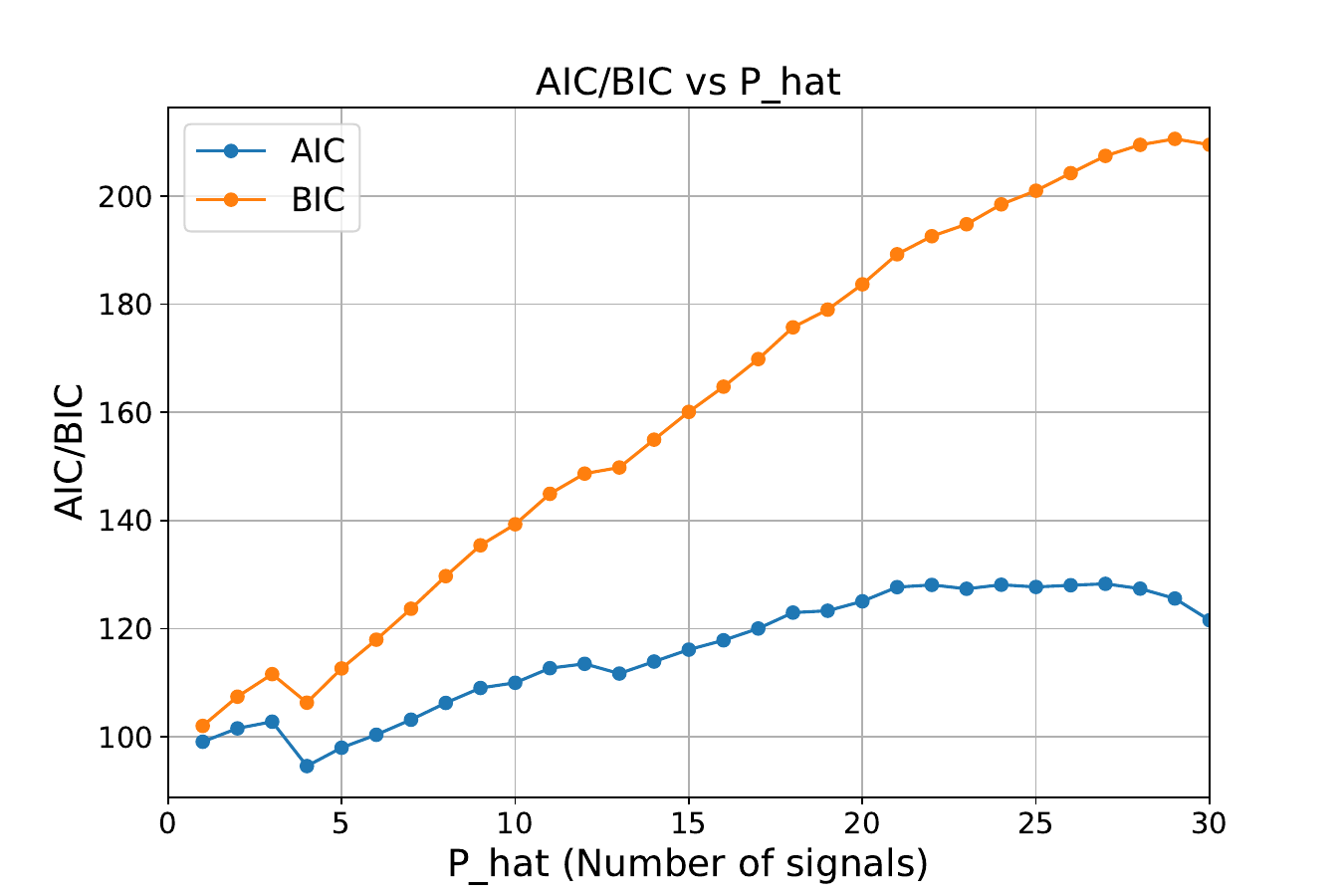}
    \caption{AIC and BIC vs $\hat P$ for $M=N=32$, $P=10$ and SNR is $20$dB.}
    \label{fig:AIC_BIC_vs_P_hat}
\end{figure}

Figure~\ref{fig:AIC_BIC_vs_P_hat} shows the values of AIC and BIC for different values of \( \hat{P} \), where \( N = M = 32 \), \( P = 10 \), and the SNR is 20~dB.  
In this example, AIC selects \( \hat{P} = 4 \), which is significantly smaller than the actual value of \( P \).  
BIC performs even worse, selecting \( \hat{P} = 1 \), which leads to a poor estimation result.  
We observed that AIC generally performs better than BIC, but it still tends to underestimate the number of paths.  
Therefore, it is preferable to consider values of \( \hat{P} \) larger than those suggested by AIC or BIC when selecting candidates for Doppler shifts.  
To this end, we employed the heuristic approach described in Section~\ref{section:MdelOrderSelection}, which often yields better estimates than those provided by AIC or BIC.
Model order selection is a central topic in machine learning, and it is expected that AI-based approaches will offer improved performance in future implementations.

\begin{figure}[t]
    \centering
    \includegraphics[width=0.85\linewidth]{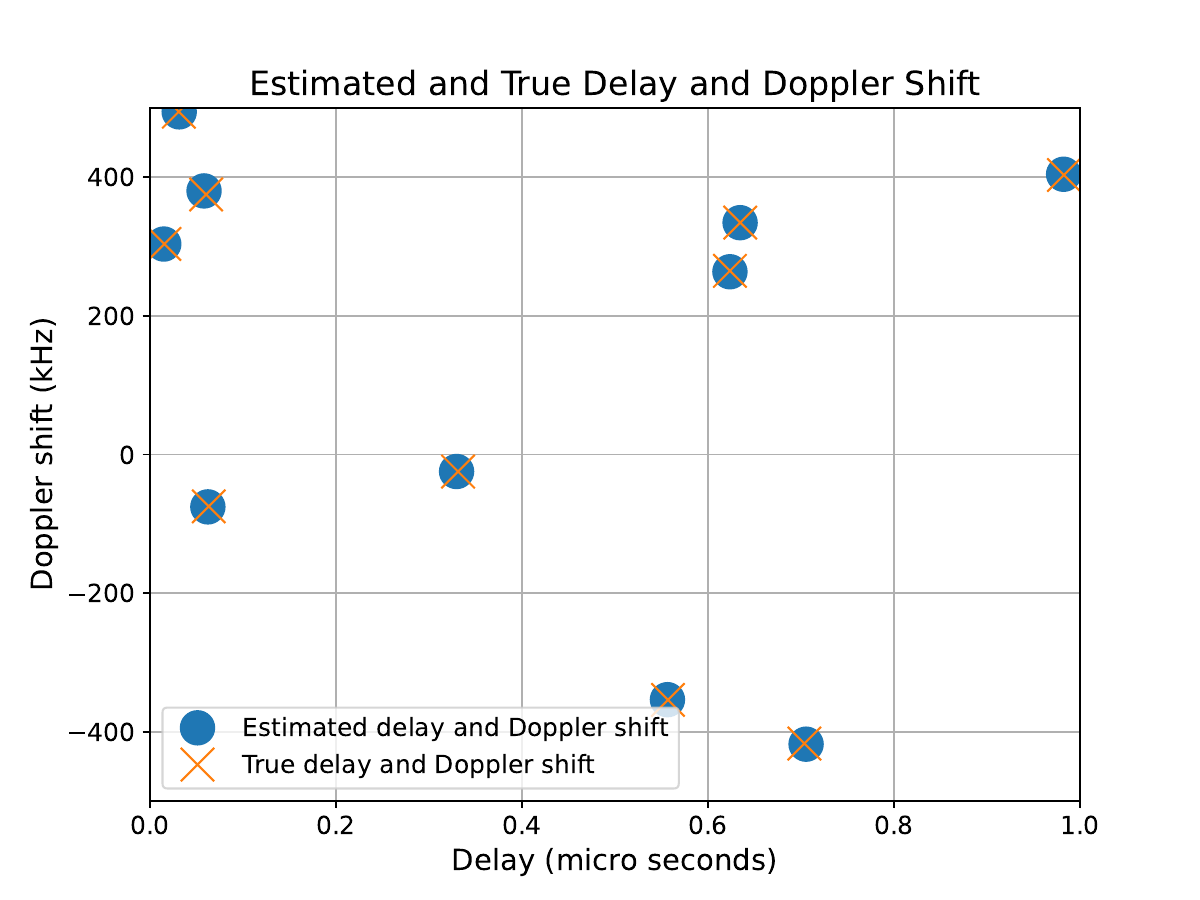}
    \caption{Estimated delay and Doppler for the same parameters in Fig.~\ref{fig:AIC_BIC_vs_P_hat}}
    \label{fig:estimated_delay_DopplerN=M=32P=10_20dB}
\end{figure}

Figure~\ref{fig:estimated_delay_DopplerN=M=32P=10_20dB} presents estimation results for \( N = M = 32 \), \( P = 10 \), and an SNR of 20~dB.  
As shown, the proposed method achieves high estimation accuracy when the number of paths is correctly specified.

\section{Concluding Remarks}
This paper discussed delay and Doppler estimation in OTFS modulation, which is essential for communication over doubly selective channels. Accurate estimation enables subsequent channel equalization. Amplitudes can also be estimated in Stage~2, which will be addressed separately.

The proposed two-stage method estimates Doppler shifts first, followed by delays, based on the structure of equation~(\ref{R=EV}). Reversing this order is also possible; in fact, paths with similar Doppler shifts but different delays can be more clearly resolved using this dual approach.

Due to the symmetry of the signal model under Fourier transform, a frequency-domain version of the method can be developed. Running both methods in parallel may improve the estimation of the number of paths and overall accuracy. This will be investigated further.

Accurate delay-Doppler estimation also supports sensing applications, including radar and joint communication-sensing systems. Although a heuristic method was used for model selection, robustness under noise remains a challenge. Integrating learning-based techniques is a promising direction.

In summary, the proposed method contributes to the integration of sensing, AI, and communication via accurate parameter estimation in OTFS systems.

\bibliographystyle{IEEEtran}
\bibliography{mybibliography.bib, gyoseki.bib}

\appendices

\section{Comparison Between the Proposed and Standard OTFS Pilot Signals}
\label{appendix:pilotsignal}
In this section, we compare our proposed pilot signal in (\ref{s(t)}) with the conventional OTFS pilot signal.  
Our proposed pilot signal can be regarded as a special case of the original OTFS signal definition:
\begin{align}
    s(t)= \sum_{m=-M/2}^{M/2-1} \sum_{n=0}^{N-1} X_\mathrm{TF}[n,m]\, g(t - mT)\, e^{j m \frac{2\pi}{T} t},
    \label{s(t)2}
\end{align}
where \( g(t) \) is a rectangular pulse of duration \( T \), and  
\( X_\mathrm{TF}[n, m] \) is the time-frequency (TF) domain representation corresponding to the delay-Doppler (DD) domain signal \( X_\mathrm{DD}[k,\ell] \).  
The TF-domain signal is obtained via the Symplectic Fourier Transform (SFT)~\cite{OTFS} as  
\begin{align}
X_\mathrm{TF}[n, m] = \frac{1}{\sqrt{NM}} \sum_{k=0}^{N-1} \sum_{\ell=0}^{M-1} X_\mathrm{DD}[k,\ell]\, e^{j 2\pi \left( \frac{kn}{N} - \frac{\ell m}{M} \right)}.
\label{SFT}
\end{align}

By substituting the DD-domain pilot in (\ref{pilot-DDdomain}) and the SFT expression in (\ref{SFT}) into (\ref{s(t)2}),  
we obtain the pilot signal given in (\ref{s(t)}).  
Note that we assume the range of \( m \) in (\ref{s(t)2}) is \( [-M/2, M/2 - 1] \).

However, in the current standard OTFS implementation—often referred to as Zak-OTFS—the delay-Doppler (DD) domain signal is directly transformed into a discrete-time domain signal via the discrete Zak transform as
\begin{align}
x_{\mathrm{TD}}[nM + \ell] = 
\frac{1}{\sqrt{N}}
\sum_{k=0}^{N-1} X_{\mathrm{DD}}[k,\ell]\, e^{j \frac{2\pi}{N} kn}.
\end{align}
The transmitted signal is, then, given by
\begin{align}
    s(t) =
    \sum_{\ell=0}^{M-1}
    \sum_{n=0}^{N-1}
    x_{\rm TD}[nM+\ell] p(t-(nM+\ell)T_s)
    \label{s(t)3} 
\end{align}

Using the form of (\ref{s(t)3}), we cannot construct exactly the same signal as (\ref{s(t)}). However, a kind of dual signal can be constructed by exchanging the time and frequency domains illustated in Figure~\ref{fig:construction_transmitted_signal}.

Suppose that $M$ is an even number.
We can select \( p(t) \) as an ideal, frequency-shifted low-pass filter such that its Fourier transform \( P(f) \)
is given by
\begin{align}
P(f) &=
\begin{cases}
1, & \text{if } -\frac{M+1}{2T} < f < \frac{M-1}{2T}, \\
0, & \text{otherwise}.
\end{cases}
\end{align}

Taking the inverse Fourier transform yields
\begin{align}
p(t) = e^{-j\frac{\pi}{T} t} \frac{\sin \left( \frac{M\pi}{T} t \right)}{\pi t}.
\end{align}

The spectral center shift is essential in the proposed method.
Such a signal is suitable for use in the two-stage Prony method with the reversed estimation order, as discussed in the Concluding Remarks.

\end{document}